\documentclass[12pt]{iopart}
\usepackage{graphicx}

\begin{document}

\def\Journal#1#2#3#4{{#1} {\bf #2}, #3 (#4)}

\def\NCA{Nuovo Cimento}
\def\NIM{Nucl. Instr. Meth.}
\def\NIMA{{Nucl. Instr. Meth.} A}
\def\NPB{{Nucl. Phys.} B}
\def\NPA{{Nucl. Phys.} A}
\def\PLB{{Phys. Lett.}  B}
\def\PRL{Phys. Rev. Lett.}
\def\PRC{{Phys. Rev.} C}
\def\PRD{{Phys. Rev.} D}
\def\ZPC{{Z. Phys.} C}
\def\JPG{{J. Phys.} G}
\def\CPC{Comput. Phys. Commun.}
\def\EPJ{{Eur. Phys. J.} C}
\def\PR{Phys. Rept.}

\title[A Novel
and Compact Muon Telescope Detector]{Perspectives of a Midrapidity
Dimuon Program at RHIC: A Novel and Compact Muon Telescope
Detector}

\author{L.~Ruan}\address{Brookhaven National Laboratory, Upton, New York 11973}\ead{ruan@bnl.gov}
\author{G.~Lin}\address{Yale University, New Haven, Connecticut 06520}
\author{Z.~Xu}\address{Brookhaven National Laboratory, Upton, New York 11973}
\author{K.~Asselta}\address{Brookhaven National Laboratory, Upton, New York 11973}
\author{H.F.~Chen}\address{University of Science \& Technology of China, Hefei 230026, China}
\author{W.~Christie}\address{Brookhaven National Laboratory, Upton, New York 11973}
\author{H.J.~Crawford}\address{University of California, Berkeley, California 94720}
\author{J.~Engelage}\address{University of California, Berkeley, California 94720}
\author{G.~Eppley}\address{Rice University, Houston, Texas 77251}
\author{T.J.~Hallman}\address{Brookhaven National Laboratory, Upton, New York 11973}
\author{C.~Li}\address{University of Science \& Technology of China, Hefei 230026, China}
\author{J.~Liu}\address{Rice University, Houston, Texas 77251}
\author{W.J.~Llope}\address{Rice University, Houston, Texas 77251}
\author{R.~Majka}\address{Yale University, New Haven, Connecticut 06520}
\author{T.~Nussbaum}\address{Rice University, Houston, Texas 77251}
\author{J.~Scheblein}\address{Brookhaven National Laboratory, Upton, New York 11973}
\author{M.~Shao}\address{University of Science \& Technology of China, Hefei 230026, China}
\author{R.~Soja}\address{Brookhaven National Laboratory, Upton, New York 11973}
\author{Y.~Sun}\address{University of Science \& Technology of China, Hefei 230026, China}
\author{Z.~Tang}\address{University of Science \& Technology of China, Hefei 230026, China}
\author{X.~Wang}\address{Tsinghua University, Beijing 100084, China}
\author{Y.~Wang}\address{Tsinghua University, Beijing 100084, China}

\date{\today}

\begin{abstract}
We propose a large-area, cost-effective Muon Telescope Detector
(MTD) at mid-rapidity for the Solenoidal Tracker at RHIC (STAR)
and for the next generation of detectors at a possible
electron-ion collider. We utilize large Multi-gap Resistive Plate
Chambers with long readout strips (long-MRPC) in the detector
design. The results from cosmic ray and beam tests show the
intrinsic timing and spatial resolution for a long-MRPC are
$60-70$ ps and $\sim1$ cm, respectively. The performance of the
prototype muon telescope detector at STAR indicates that muon
identification at a transverse momentum of a few GeV/$c$ can be
achieved by combining information from track matching with the
MTD, ionization energy loss in the Time Projection Chamber, and
time-of-flight measurements. A primary muon over secondary muon
ratio of better than 1/3 can be achieved. This provides a
promising device for future quarkonium programs and primordial
dilepton measurements at RHIC. Simulations of the muon efficiency,
the signal-to-background ratio of $J/\psi$, the separation of
$\Upsilon$ 1S from 2S+3S states, and the electron-muon correlation
from charm pair production in the RHIC environment are presented.

\end{abstract}
\section{Introduction}\label{intro}
Data taken over the last several years have demonstrated that RHIC
has created dense and rapidly thermalizing matter characterized
by: 1) initial energy densities far above the critical values
predicted by lattice QCD for formation of a Quark-Gluon Plasma
(QGP); 2) opacity to jets; and 3) nearly ideal fluid flow, which
is marked by constituent interactions of very short mean free
path, established most probably at a stage preceding hadron
formation~\cite{starwhitepaper}. The next objective at RHIC is to
study properties of this partonic matter in detail in terms of
color degrees of freedom and the equation of state. For example,
due to color screening, different quarkonium states will
dissociate and the dissociation temperatures will be different due
to different binding energies. The precise measurement of
transverse momentum distributions of quarkonia at different
centralities, collision systems, and energies will serve as a
thermometer of QGP. A large-area Muon Telescope Detector (MTD) at
mid-rapidity for RHIC collisions will be crucial for advancing our
knowledge of QGP properties. It will directly address many of the
open questions and long-term goals proposed in the RHIC white
papers~\cite{starwhitepaper,rhicwhitepaper}. Since muons do not
participate in strong interactions, they provide penetrating
probes for the strongly-interacting QGP. A large area detector
identifying muons with momentum of a few GeV/$c$ at mid-rapidity
allows for the detection of di-muon pairs from QGP thermal
radiation, quarkonia, light vector mesons, possible correlations
of quarks and gluons as resonances in QGP, and Drell-Yan
production, as well as the measurement of heavy flavor hadrons
through their semi-leptonic decays into single
muons~\cite{dilepton}. Some of these topics also can be studied
using electrons or photons or a combination of both. However,
electrons and photons have large backgrounds from hadron decays,
$\pi^{0}$ and $\eta$ Dalitz decays, and gamma conversions in the
detector material. These backgrounds prevent effective triggering
in central nucleus-nucleus collisions at mid-rapidity at the
Solenoidal Tracker at RHIC (STAR). In addition, electron-muon
correlation can be used to distinguish between lepton pair
production and heavy quark decays ($c+\bar{c}\rightarrow
e+\mu(e)$, $B \rightarrow e(\mu)+ c \rightarrow e+\mu(e)$). In
addition, muons are less affected by Bremsstrahlung radiation
energy loss in the detector materials than electrons, thus
providing excellent mass resolution of vector mesons and
quarkonia. This is essential for separating the ground state (1S)
of $\Upsilon$ from its excited states (2S+3S). They are predicted
to melt at very different temperatures.

Conventional muon detectors rely heavily on tracking stations,
while this new detector proposes to use  $\!<\!100$ ps timing and
$\sim\!1$ cm spatial  resolution to identify muons with momentum
of a few GeV/$c$~\cite{MTDLDRD}. Multi-gap resistive plate chamber
technology with large modules, long strips and double-ended
readout (long-MRPC) was used for MTD prototype detectors. Similar
technology but with small pads has been constructed and installed
at STAR as a Time-of-Flight Detector (TOF)~\cite{startofproposal}.
In this report we present the conceptual design for the STAR MTD,
the R$\&$D results including the intrinsic timing and spatial
resolution of long-MRPC, and the MTD prototype performance at
STAR. The muon identification capability and hadron background
rejection are reported. Future perspectives for physics programs
utilizing such detectors are discussed.

\section{Simulation}\label{simu}
\begin{figure}
\begin{center}
\includegraphics[keepaspectratio,scale=0.7]{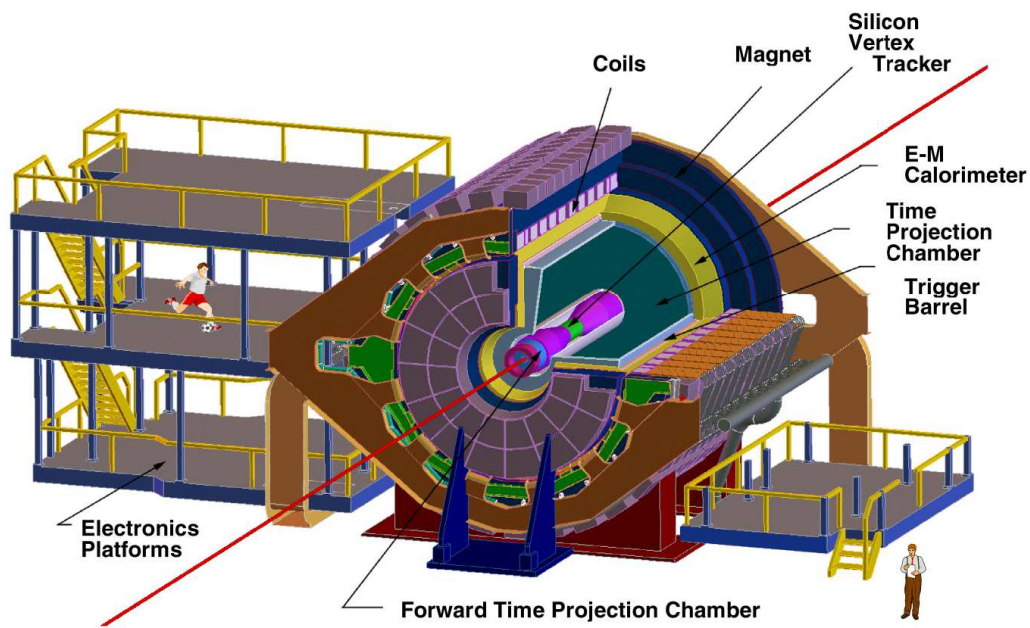}
\caption{ Perspective view of the STAR detector, with a cutaway
for viewing inner detector systems. The trigger barrel, shown in
this figure, will be fully replaced by the TOF detector in year
2009. Figure is taken from~\cite{star}.} \label{stardetector}
\end{center}
\end{figure}

The STAR detector was used for these studies~\cite{star}. The
detector layout is shown in Fig.~\ref{stardetector}. The main
tracking device is the Time Projection Chamber
(TPC)~\cite{startpc}, whose inner and outer field cages are
located at radial distances of 50 and 200 cm respectively from the
beam axis. The TPC is 4 meters long and it covers a pseudorapidity
range $|\eta|\!<\!1.8$ and 2$\pi$ in azimuth. The ionization
energy loss ($dE/dx$) is used for particle
identification~\cite{pidNIMA,bichsel,pidpp08}. A TOF detector
based on Multi-gap Resistive Plate Chambers (MRPC)~\cite{startof}
will be fully installed in STAR in 2009, covering $2\pi$ in
azimuth and $-1\!<\!\eta\!<\!1$ in pseudorapidity at a radius of
$\sim220$ cm.  It will extend particle identification up to
$p_{T}\sim3$ GeV/$c$ for $p$ and $\bar{p}$~\cite{tofPID}. The full
barrel electromagnetic calorimeter (BEMC) is installed outside the
TOF radius and uniformly covers $-1\!<\!\eta\!<\!1$ in
pseudorapidity and $2\pi$ in azimuth~\cite{starbemc}. The TPC is
centered in a solenoidal magnetic field provided by the
surrounding magnetic coils. The return flux path for the field is
provided by the magnet steel~\cite{starmagnet}, which is roughly
cylindrical in geometry and consists of 30 flux return bars, four
end rings and two poletips. The 6.85 m long flux return bars are
trapezoidal in cross-section and 60 cm thick with a 363 cm outer
radius. The width at the outer radius of the return bar is 57 cm.

\begin{figure}
\begin{center}
\includegraphics[keepaspectratio,scale=0.7]{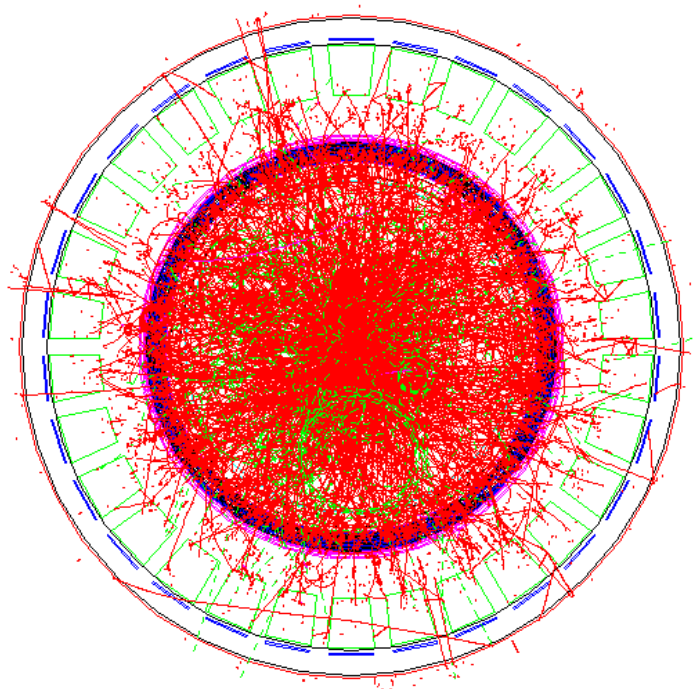}
\caption{ A full HIJING central Au+Au collisions simulated in
STAR. The tracks in the TPC are projected in the plane transverse
to the beam line. } \label{figure1} \end{center}
\end{figure}

The simulation of a full HIJING central Au+Au collision is shown
in Fig.~\ref{figure1}, using STAR geometry with a configuration of
all the detectors and a complete material budget. We simulated a
muon-detector (in blue) covering the full return bars (in green)
within $|\eta|\!<\!0.8$ and left the gaps in-between return bars
uncovered. This detector acceptance corresponds to 56\% of $2\pi$
in azimuth. In Fig.~\ref{figure1}, it can be seen that most of the
particles are stopped before the BEMC and the few escaping
particles (primary or secondary) mainly come through the gaps
between the return bars.

\begin{figure}
\begin{center}
\includegraphics[keepaspectratio,scale=0.7]{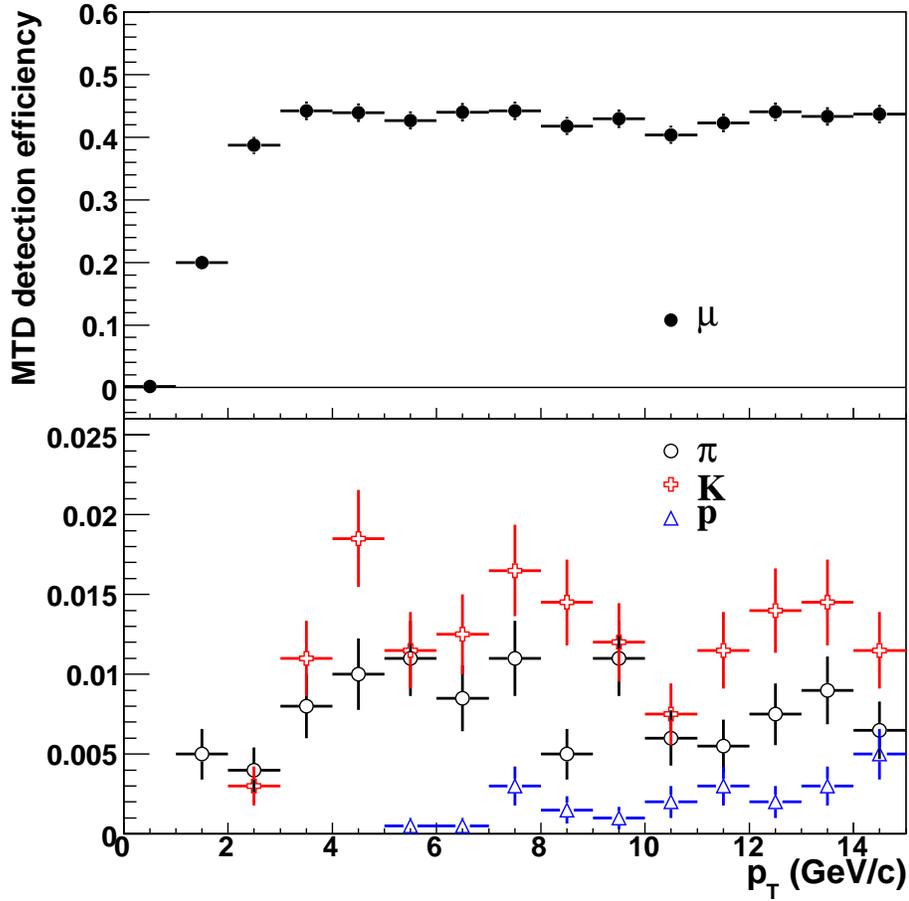}
\caption{Efficiency for muons (top panel) and misidentification
probabilities of pions, kaons and protons (bottom panel). The
inefficiency for the muon is mainly due to the acceptance of the
MTD (56\%).} \label{figeff} \end{center}
\end{figure}

Further simulation~\cite{DNP06} with STAR geometry indicates that
for a muon with $p_{T}\!>\!2$ GeV/$c$ generated in the center of
the TPC, the detection efficiency of the MTD, including acceptance
effects, is about 40-50\%, while for a pion track, the efficiency
is 0.5-1\%. In this simulation, the charged tracks reconstructed
in the TPC are extrapolated to the MTD, and required to match the
hit position as well as the time-of-flight from MTD measurements.
We require the distance between the MTD hit and the projected
point from the track to be within 4 cm and the timing difference
to be less than 400 ps. It should be noted that in this study, a
simple helix extrapolation of the track to the MTD was employed
while in reality, the magnetic flux return bar has a magnetic
field of 1.2 Telsa while the TPC has a 0.5 Telsa field. Pion and
kaon misidentifications come mainly from their decay into muons
before the absorption by the return bar. These daughter muons then
penetrate the return bars to reach the MTD. It is also evident
from the simulations that protons are reduced by a much larger
factor than pions and kaons. Shown in Fig.~\ref{figeff} is the
combined acceptance and efficiency for primary muons, pions, kaons
and protons. It indicates that the MTD can detect muons of
$p_T{}^{>}_{\sim}2$ GeV/$c$ at a level of 45\% while less than 1\%
of overall hadrons are accepted taking into account the relative
yields of pions, kaons and protons. This illustrates that the MTD
detector can reject hadrons effectively. A rejection factor of
50-100 can be obtained based on the simulations shown in
Fig.~\ref{figeff} while maintaining an efficient trigger ($>80\%$)
for prompt muons. Further hadron rejection can be achieved with
tracking in the TPC, $dE/dx$ in the TPC and hit matching in the
TOF. These have been studied in detail with a prototype MTD tray
in STAR. The results will be discussed in the next few sections.
With the muon efficiency shown above and the measured $J/\psi$
$p_{T}$ distributions~\cite{highptjpsi}, we obtained $J/\psi$
efficiency versus $p_T$, as shown in Fig.~\ref{figjpsieff}.

\section{The R$\&$D results for the MTD}\label{RandDresults}

We utilize long-MRPCs for the detector design. Each long-MRPC
module consists of two stacks of resistive glass plates resulting
in a combined total of ten uniform gas gaps with gap widths of 250
$\mu$m. High voltage is applied to electrodes on the outer
surfaces of the outer plates of each stack. A charged particle
traversing a module generates avalanches in the gas gaps which are
read out by six copper pickup strips with strip dimensions of
$870\times25$ $\mathrm{mm}^{2}$. The MRPC modules were operated at
12.6 kV with a mixture of 95\% $C_{2}H_{2}F_{4}$ and 5\%
iso-butane at 1 atmosphere. In the high voltage range
$\!12.5\!<\!HV\!<\!13.0$ kV, the efficiency is above 95\% and
timing resolution is about 60-70 ps in the cosmic ray and beam
tests. The spatial resolution of the long-MRPC along the long
strip is about 0.6-1 cm in the same tests. This satisfies the
needs for a large-area muon detector. The details of the long-MRPC
construction and its performance in the cosmic ray and beam tests
can be found in this paper~\cite{MTDNIMA}.
\begin{figure}
\begin{center}
\includegraphics[keepaspectratio,angle=90,scale=0.4]{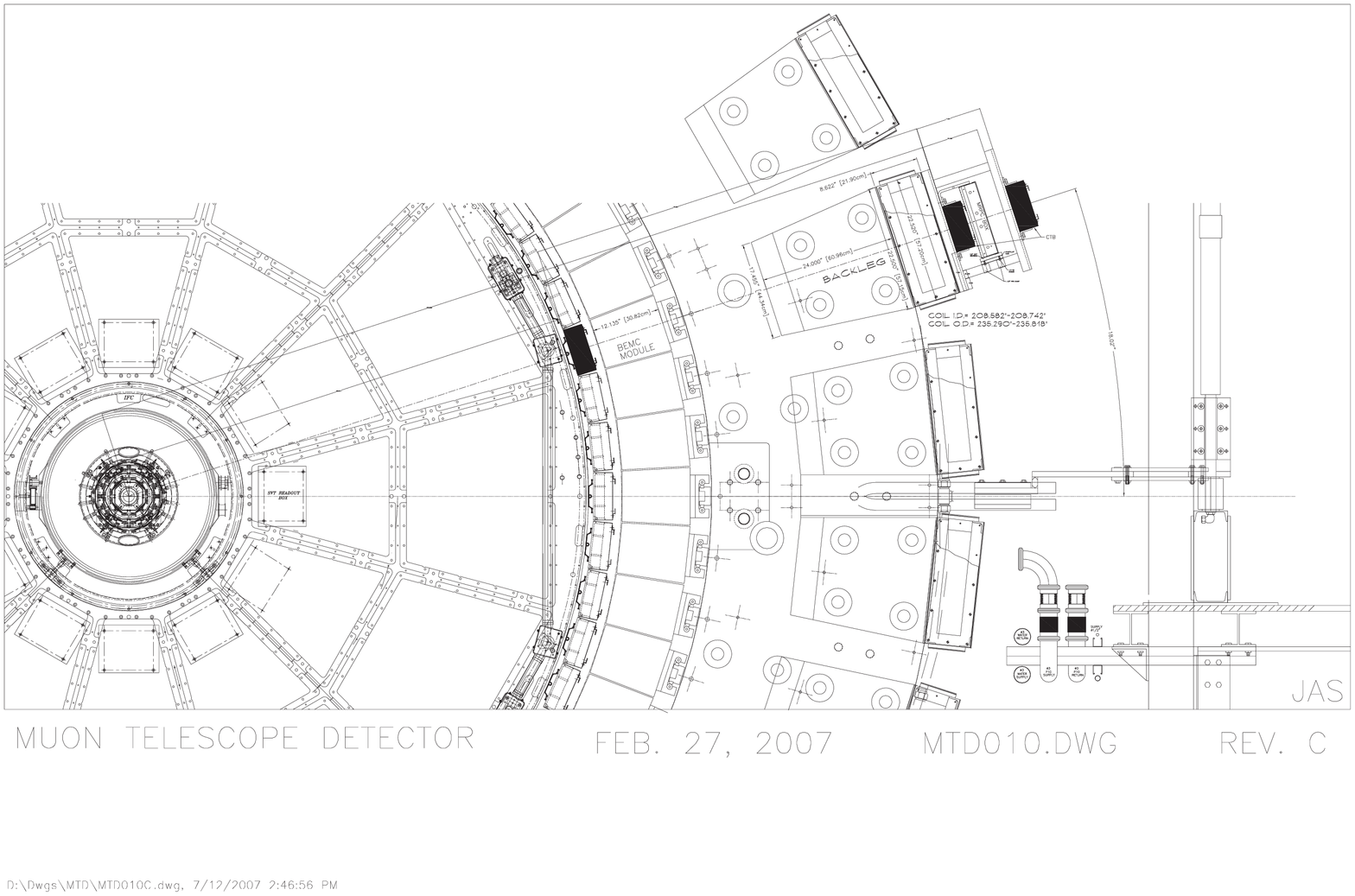}
\caption{ Position of the MTD prototype at STAR. The MTD was
installed at 400 cm radially from interaction point in
2006--2008.} \label{MTDdrawing} \end{center}
\end{figure}

\section{Performances of a Prototype MTD}\label{test}

\begin{table}{}
\caption{\label{Tab:A}Trigger rates, sampled luminosities,
recorded events and matched hits with TPC tracks at $p_T\!>$ 1.5
GeV/$c$ in Au+Au, d+Au and p+p collisions in year 2007 and 2008
with the prototype MTD.}
\begin{scriptsize}
{\centering {\begin{tabular}{|c|c|c|c|c|c|} \hline \hline beam
species & interaction rate (Hz)& trigger rate (Hz) & sampled
luminosity & recorded events (M) & matched hits \\ \hline Au+Au &
20 k &0.5--2 & 270 $\mu b^{-1}$ & 0.31 & 7 k \\ \hline d+Au & 100
k &0.5--2 & 29 $nb^{-1}$ &1.6 & 78 k \\ \hline p+p & 300 k &
0.5--2 & 404 $nb^{-1}$ &0.56 & 8 k \\ \hline \hline
   \end{tabular}
 }
\par}
\end{scriptsize}

\end{table}

A prototype of the MTD was installed and covered $\pi/60$ in
azimuth and $-0.25\!<\!\eta\!<\!0.25$ in pseudorapidity at a
radius of $\sim400$ cm during the 2007 run in 200 GeV Au+Au
collisions. During the 2008 run in 200 GeV d+Au and p+p
collisions, the prototype was re-located and covered
$-0.5\!<\!\eta\!<\!0$ in pseudorapidity at the same radius. It
contained two long-MRPC modules. The prototype was placed outside
of the return bars which serve as a hadron absorber, each of which
amounts to 5 interaction lengths. A mechanical drawing of the
location and geometry of the MTD in STAR is shown in
Fig.~\ref{MTDdrawing}. The analog signals~\cite{mtdelectronics}
were sent to trigger digitizer boards specially developed for use
in the STAR trigger system ~\cite{startrigger} for signal
amplitude and timing information read-out. Each digitizer board
consists of an 8 bit analog-to-digital convertor (ADC).  In
addition, discriminator outputs from the digitizer boards with
thresholds set to 30 mV were routed to a time-to-analog convertor
(TAC), which was gated by the RHIC accelerator clock. The TAC
signals were sent to a different channel of the digitizer board.
Both ADC and TAC values were routed through the trigger
distribution and processing system for an initial (level 0)
trigger decision~\cite{startrigger}. The level 0 (L0) trigger is
generated in less than 1.5 us from all the fast detector
information available to trigger and is used to start the
digitization process on slower detector systems (e.g. TPC). We
have not required time difference or any other matching algorithms
for a trigger decision as discussed in the simulation. A valid hit
was required to have a non-zero and non-overflow ADC and TAC value
within the 8-bit digitizer board range. The prototype successfully
triggered the data acquisition system by requiring a valid hit in
at least one strip of the long-MRPCs. Table~\ref{Tab:A} lists
trigger rates, sampled luminosities and recorded events in Au+Au,
d+Au and p+p collisions in year 2007 and 2008 with the prototype
MTD. For the year 2009 run, we have installed another prototype
tray, which is equipped with the same electronics as is used in
the TOF system at STAR~\cite{startofproposal} to further improve
the timing resolution of the MTD at STAR. The TOF electronics has
a time bin width (25 ps) that is small compared to the detector
resolution, and the degrading effects of the long cables are
removed as the digitization will be done on-board the detector.
The addition of the TOF-based electronics is expected to allow a
significant improvement to the MTD's timing measurement in the
off-line analysis.

\begin{figure}\begin{center}
\includegraphics[keepaspectratio,scale=0.7]{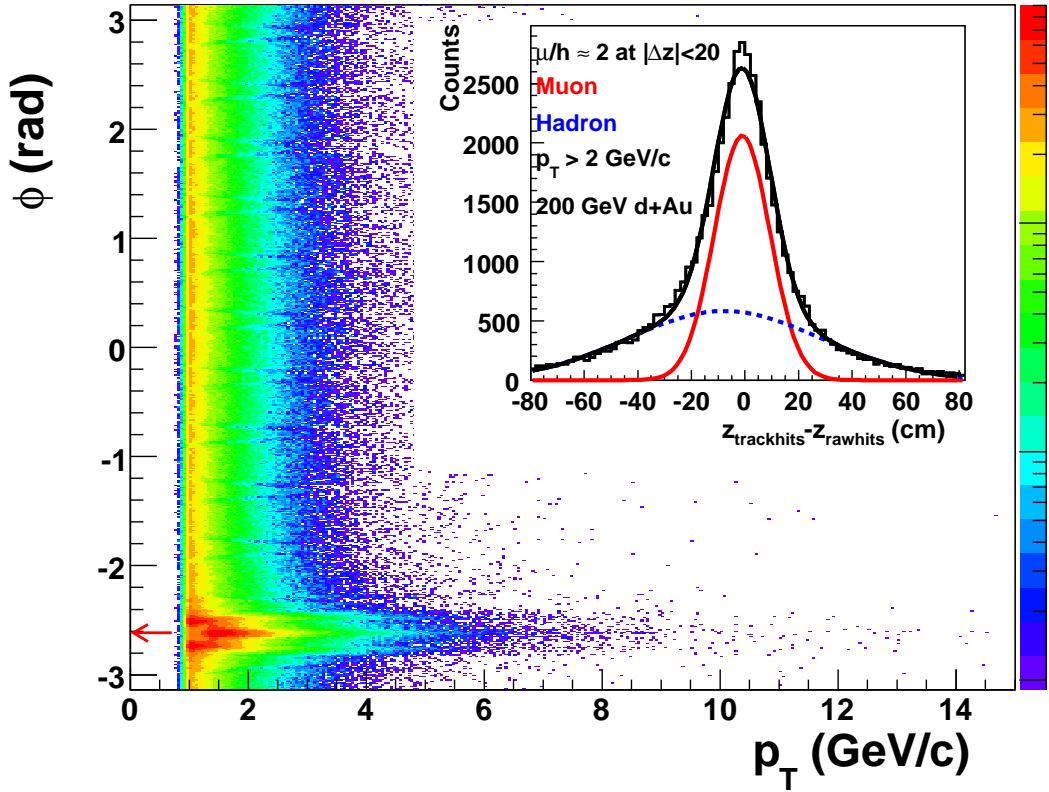}
\caption{ Azimuthal angle distribution of particles versus $p_{T}$
in d+Au collisions, extrapolated from the TPC to a radius of 400
cm. The arrow denotes the azimuthal location of the prototype MTD
tray. The insert shows $\Delta$z distribution. $z_{trackhits}$ is
derived from extrapolation of the tracks of the TPC to the MTD
location. $z_{rawhits}$ is measured by the time difference from
two-end readout of the hit strip along the long strip of the
long-MRPC. } \label{figure3}
\end{center}\end{figure}

Shown in Fig.~\ref{figure3} is the azimuthal angle distribution of
particles from the TPC extrapolated to a radius of 400 cm in
triggered d+Au collisions as a function of the transverse momentum
$p_{T}$. The peak shows an enhancement of particle yield at the
angle where the MTD is positioned. This shows that offline
tracking of particles from the TPC was able to match hits from the
long-MRPC. The tracks of the TPC were extrapolated to the MTD
location, resulting in position information from tracking. The
time difference from two-end readout of the hit strip provides a
position measurement along the long strip of the long-MRPC. The
MTD is placed behind the backleg steel of the magnet thus multiple
hits are not an issue. The difference of these two position values
in the z direction ($\Delta$z) is shown in the insert of
Fig.~\ref{figure3}, where the z direction is the beam direction. A
double Gaussian function was used to fit the distribution. The
$\sigma$ of the narrow Gaussian was found to be $\sim\!10$ cm by
selecting tracks of $p_{T}\!>\!2$ GeV/$c$ while the other Gaussian
is significantly broader. The ratio of the particle yields in the
narrow Gaussian to those in the broad Gaussian, within our match
window at $\!|\Delta z\!| \!<\!20$ cm (the narrow-to-broad ratio)
is $\sim$2. Table~\ref{Tab:A} lists the total matched hits within
the match window $\!|\Delta z\!| \!<\!20$ cm at $p_T\!>$ 1.5
GeV/$c$ in Au+Au, d+Au and p+p collisions with the prototype MTD.
From GEANT simulation, muons of $p_{T}\sim\!2.5$ GeV/$c$ generated
at the center of the TPC result in a Gaussian distribution with a
sigma of 9 cm in the z direction in the MTD barrel, after
traversing the detector materials from the TPC center to the MTD.
The simulation also indicates that pions will result in a much
broader distribution due to the strong interaction.

\section{Muon identification capability and hadron rejection}\label{primarymuon}

At $p_T\!>\!2$ GeV/$c$, the $dE/dx$ of a muon track is
$\sim\!3-\!4\%$ higher than that of a pion track~\cite{PDG}. The
resolution of the $dE/dx$ in the TPC is
$\sim\!8\%$~\cite{startpc}, therefore, a half sigma difference in
$dE/dx$ is expected between a muon and a pion track. About a
two-sigma difference is expected between a muon and a kaon track.
The energy loss due to multiple scattering in the return bar will
cause an energy loss of $\sim$1.2 GeV for muons at
$2\!<p_{T}\!<\!10$ GeV/$c$ while the velocity change is
negligible. For hadrons, the strong interaction will lead to a
hadronic shower.  This results in a significant energy loss and a
change of velocity, and therefore a later arrival at the MTD. By
selecting high velocity and large $dE/dx$ particles, the hadron
background was observed to be significantly reduced and the
narrow-to-broad ratio was found significantly enhanced. This is
consistent with the expectation that the narrow Gaussian is
dominated by muons and the broad Gaussian is dominated by hadrons.
Within the match window $\!|\Delta z\!| \!<\!20$ cm, muon purity
can be achieved to greater than 80\% through the combined
information of track matching, $dE/dx$, and velocity.

The average long-MRPC timing resolution for the two modules used
in this analysis was measured to be $\sim\!300$ ps in Au+Au
collisions. The ``start'' time was provided by two identical
vertex position detectors (VPDs), each 5.4 meter away from the
nominal collision point along the beamline~\cite{upVPD}. After
subtracting the start timing resolution ($\sim\!160$ ps) and
detector material effect contribution ($\sim\!100$ ps at
$p_T\!=\!2.5$ GeV/$c$), the timing resolution ($200-300$ ps) from
the MTD was found to be worse than those from cosmic and beam
tests. This is understood by the fact that the trigger read-out
electronics was not designed for precise time measurement. With
the proposed full scale detector, we will use TOF electronics. TOF
electronics has a timing resolution of 25 ps. It has proved to be
a low-cost, reliable system for recording timing information,
reading it out, and sending it to STAR DAQ. The electronics boards
used for TOF can be used for the MTD efficiently, without any
design changes.

\begin{figure}\begin{center}
\includegraphics[keepaspectratio,scale=0.7]{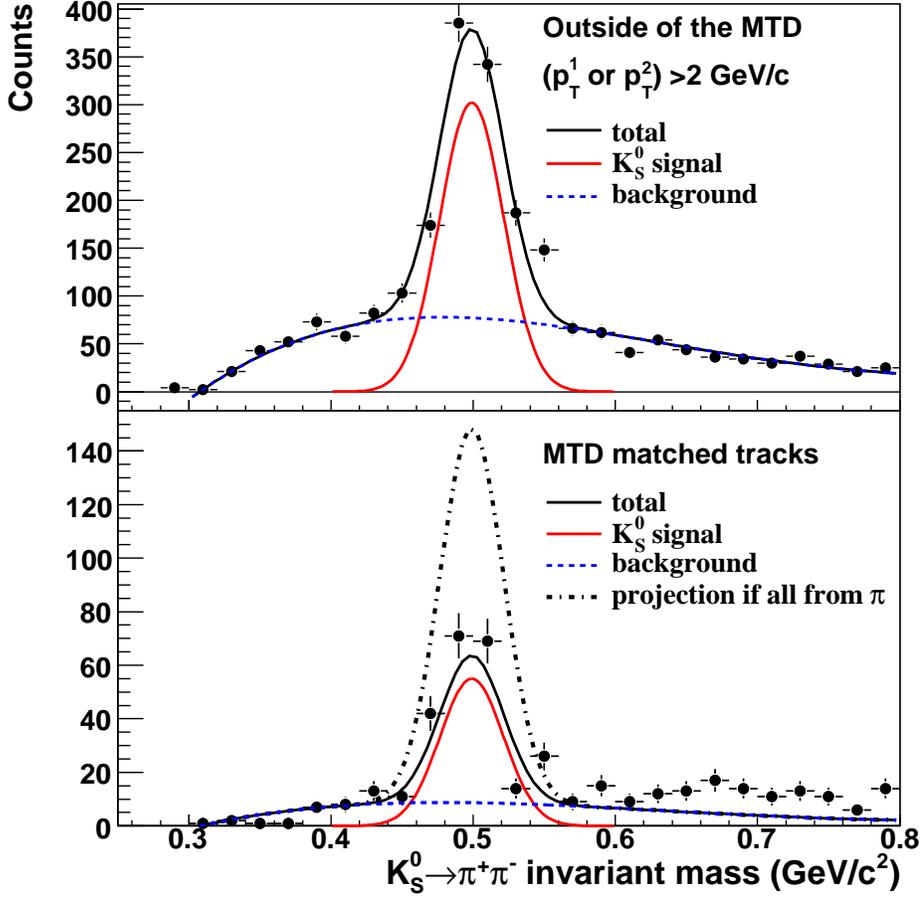}
\caption{ Reconstructed $K^{0}_{S}$ which are not associated with
MTD triggered particles (top panel); reconstructed $K^{0}_{S}$
with at least one of the daughter pions associated with the MTD
triggered particle. The dot-dashed line depicts the $K^{0}_{S}$
yields from a projection assuming that the particles associated
with MTD hits are all from pion decays (bottom panel). }
\label{figKs}
\end{center}\end{figure}

To further assess the pion contamination, we identify the pion
tracks which come from the $K_{S}^{0}$ weak decay through the
hadronic decay channel $K_{S}^{0}\rightarrow\pi^{+}\pi^{-}$ to
measure the fraction of muon candidates from pion decays.
$K_{S}^{0}\rightarrow\pi^{+}\pi^{-}$ was reconstructed through
decay V0 topology. Shown in Fig.~\ref{figKs} (top panel) is the
raw $K^{0}_{S}$ from pion pairs reconstructed in other azimuthal
angles away from the MTD while in the bottom panel is the raw
$K^{0}_{S}$ yield with at least one of the daughter pions
associated with a MTD hit. The dot-dashed line depicts the
$K^{0}_{S}$ yields from a projection assuming that the particles
associated with MTD hits are all from pion decays. We found the
secondary muons from pion decay contributed 30-40\% to the total
muon candidates. The remaining background contaminating the prompt
muon candidates was secondary muons from kaon decay. This can be
investigated using the difference of TPC track $dE/dx$ for kaons
and muons. By cutting on a high $dE/dx$ value of
$n\sigma_{\pi}\!>\!0$, where $n\sigma_{\pi}$ is the measured
$dE/dx$ normalized by the value of expected pion
$dE/dx$~\cite{pidNIMA}, we reject part of the kaon secondary decay
products. Furthermore, from p+p MTD triggered events taken in year
2008, we found that the inclusive muon yields can be decreased by
a factor of 2 by requiring a coincidence with a valid hit in the
TOF system in the same TPC sector.

\begin{table}{}
\caption{\label{mip}The fraction of TPC tracks which match with a
non-zero EMC energy ($>$ MIP) from non-MTD sectors (A), and the
fraction of TPC tracks which match with a non-zero EMC tower
energy ($>$ MIP) and a MTD hit (B). A non-zero EMC energy is
defined to be $>0.75$ GeV above pedestal.} {\centering
{\begin{tabular}{|c|c|c|} \hline\hline $p_T$ (GeV/$c$) & EMC $\&$
non-MTD (A) & EMC $\&$ MTD (B)
\\ \hline 1.5--2 &($2.21\pm0.03$)\% &($0.18\pm0.02$)\%\\ \hline 2--3 & ($3.67\pm0.07$)\% &($0.30\pm0.04$)\%
\\ \hline 3--4 & ($5.89\pm0.22$)\% & ($0.34\pm0.10$)\% \\ \hline 4--6&($7.92\pm0.50$)\% & ($0.65\pm0.24$)\%
\\ \hline \hline
   \end{tabular}
 }
\par}
\end{table}

We not only use several detectors (TPC, TOF and MTD) to improve
the muon identification, but also use additional detectors to
check the performance of hadron rejection. One of the methods is
to use hadronic showers in the BEMC to study the fraction of
inclusive muon candidates from hadron showers starting at the EMC
and penetrating the absorbers. Only a few percent of hadrons
produce showers and deposit energy above the minimum ionization
energy (MIP) in the BEMC. Table~\ref{mip} shows the fraction of
hadrons with energy deposited in the BEMC greater than 0.75 GeV
with and without any MTD hit associated. There is a factor of
$\ge10$ reduction of hadronic shower in the BEMC when we require
muon candidacy in the MTD. Since the electron yield is about 0.1\%
of the inclusive hadron yield in year 2008 in 200 GeV p+p
collisions at STAR~\cite{FuSQM08}, it is reasonable to assume that
all the BEMC showers with a TPC track are hadrons when a MTD hit
is not associated. This means that the contribution from hadrons
was reduced from $\simeq100\%$ to $\le10\%$ while that from muons
(both prompt and from hadron decays) increased to $\ge90\%$ of the
remaining sample. By comparing the difference between the BEMC
showers with and without the MTD hit requirement, we conclude that
particles which arrive at the BEMC when there is a valid MTD hit
in coincidence are indeed mostly dominated by muons. However, the
muons are primarily from pion and kaon decays in the first 2
meters of the STAR detector.

\begin{figure}\begin{center}
\includegraphics[keepaspectratio,scale=0.7]{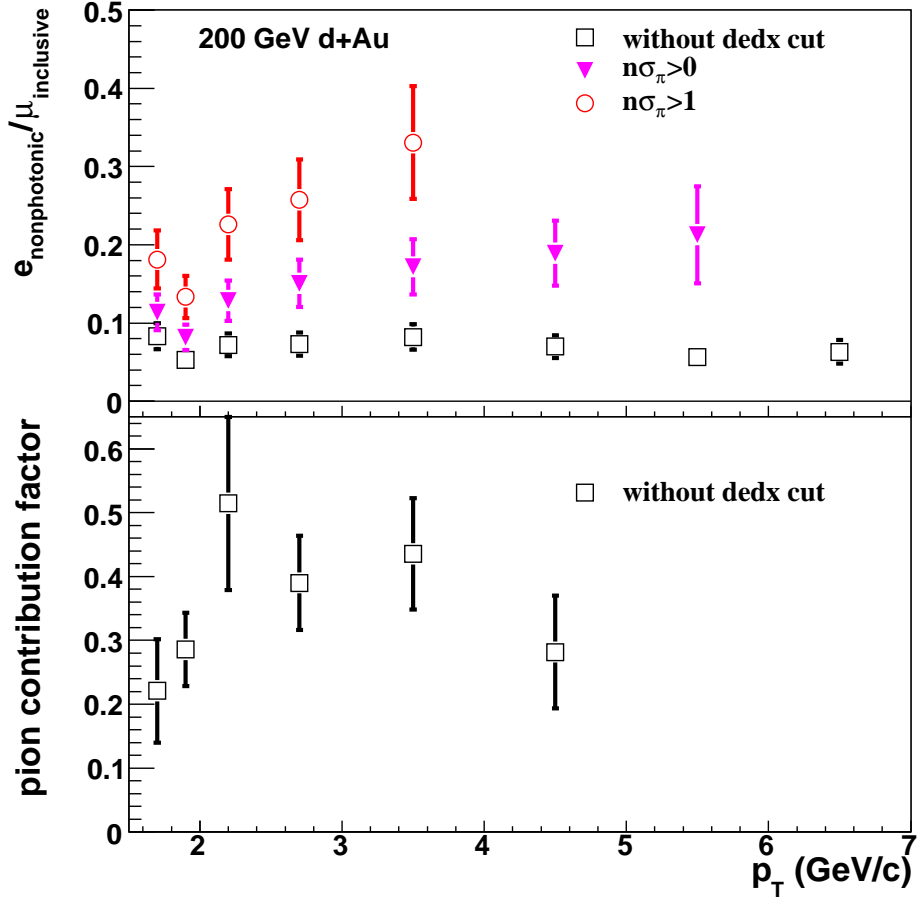}
\caption{ Estimation as a function of $p_T$ of the fraction of
detected muons which are prompt by comparison of the total
inclusive muon yield to the measured yield of non-photonic
electrons (electrons not resulting from photon conversion, hadron
decays or Dalitz decays). The estimate is made with and without
cuts on the ionization energy loss (relative to that expected for
a pion) in the TPC for the track projected to the position of the
MTD.  Since the dE/dx for a muon in the relevant momentum range is
higher than that for a pion, increasing the cut on $n\sigma_{\pi}$
results in a higher fraction of prompt muons (top panel); the
contribution from pion decays to the measured inclusive muon
signal as a function of $p_T$, determined by measuring the MTD
response when a pion from a $K^{0}_{S}$ decay enters its
acceptance (bottom panel).} \label{mu2pi}
\end{center}\end{figure}

Since the prompt muons are predominantly from charm semi-leptonic
decay, we compare our muon yields to non-photonic electron yields
in the same $p_T$ bin to assess the muon contamination from pion
and kaon decays. The raw yields of muons were obtained from
Gaussian fits to the distributions of $\Delta z$. The narrow
component is attributed to the muon. The systematic uncertainty on
the muon raw yields from different fit ranges and parameter
constraints is about 20\%. The acceptance and efficiency were
studied in simulations and corrected for the yields. The ratio of
the raw yields $d^2N/(2{\pi}p_Tdp_Tdy)$ of inclusive muons
($(\mu^{+}+\mu^{-})/2$) to the non-photonic electron yields in 200
GeV d+Au collisions at mid-rapidity are shown in Fig.~\ref{mu2pi}
for different muon selection criteria. Also shown in
Fig.~\ref{mu2pi} is the pion related contribution to the inclusive
muon yield as a function of $p_T$, evaluated by reconstructing
$K^{0}_{S}$ with at least one of the daughters associated with the
MTD hit, as described above in detail. The non-photonic electron
invariant yields are from STAR publications~\cite{starelectron}.
Compared to the non-photonic electron yields, the primary muons
contributed 6-10\% to the inclusive muons if only MTD hit
association was applied. Table~\ref{Tab:D} lists the primary muon
selection efficiency and the ratio of primary muon over secondary
muon from pion and kaon decay (S/B) under different selection
criteria on the $dE/dx$. By selecting on a high $dE/dx$ value of
$n\sigma_{\pi}\!>\!1$, the S/B ratio increased by a factor of 3,
as shown in Table~\ref{Tab:D}.

\begin{table}{}
\caption{\label{Tab:D}The signal-to-background ratio and primary
muon selection efficiency under different conditions in 200 GeV
d+Au collisions. The S/B ratio is $p_T$ dependent.} {\centering
{\begin{tabular}{|c|c|c|} \hline \hline selection criterion &
 primary muon efficiency & the S/B ratio\\
 without cut & 100\% &1/15--1/9\\
$n\sigma_{\pi}\!>\!0$ & 61\% & 1/10--1/6\\
$n\sigma_{\pi}\!>\!1$ & 26\% & 1/6--1/3\\
 \hline \hline
   \end{tabular}
 }
\par}
\end{table}

\section{Trigger Capability}

The full muon telescope detector will cover 56\% of the TPC
acceptance at $\!|\eta\!|\!<$0.8. If the similar design as for the
prototype tray is used, there will be 540 long-MRPC modules, 2160
read-out strips, and 4320 read-out channels, which is a factor of
5-6 less than what have been used for the TOF at STAR. By using
the same electronics design, the expense for the full coverage of
MTD system will be very cost effective, and will likely be similar
in scope to the smaller RHIC II upgrades.

\begin{table}{}
\caption{\label{Tab:E}The trigger rate reduction by requiring a
single hit or double hits in the MTD for different collision
systems at 200 GeV with full MTD coverage. L0: single-hit L0 rate
output from MTD for a given collision event; L2 with TOF:
single-hit rate reduction at L2 trigger with requirement of a TOF
hit at the vicinity of the path of a muon candidate (the rate in
p+p collisions is measured; the rate in central Au+Au collisions
is estimated only); RHIC II Lumi.: RHIC II luminosity in terms of
collision rate; dimuon L2 rate: dimuon trigger rate at L2 in Hz at
RHIC II. The numbers in the parenthes are from simulations with a
tighter timing cut assuming the electronics can provide $\sim100$
ps resolution.}

{\centering {\begin{tabular}{|c|c|c|c|c|c} \hline \hline collision
system & L0 & L2 with TOF & RHIC II Lumi. (Hz)&
dimuon L2 rate (Hz)\\ \hline minbias Au+Au & 0.2 &--& &\\
0-5\% Au+Au& 0.5 (0.2) &$\sim0.3$& 100 k & $\sim500$ (100)\\ 60-80\% Au+Au & 0.03 &--&&\\
d+Au & 0.02 & --& &\\ p+p & 0.01 & 0.003& 2 M & $\le10$\\ \hline
\hline
   \end{tabular}
 }
\par}
\end{table}

\begin{figure}\begin{center}
\includegraphics[keepaspectratio,scale=0.8]{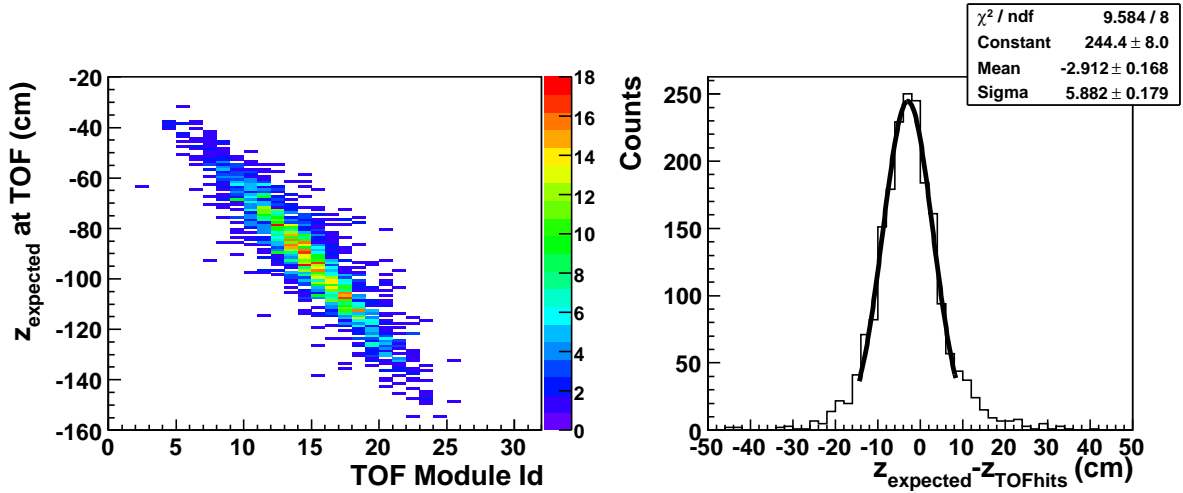}
\caption{ The expected z position at the TOF system versus the
module ID with a valid hit in 200 GeV p+p collisions in year 2008.
The modules in the TOF system are sequentially positioned along
the z direction (left panel); the distributions of the difference
between the expected position and the measured position along the
z direction from the TOF system (right panel). The expected
position ($z_{expected}$) is obtained From the extrapolation of a
valid MTD hit and the time difference measured between the VPDs on
the east and west sides of the interaction point.}
\label{ztofcorr}
\end{center}\end{figure}

Based on the MTD trigger rate from the prototype tray in Au+Au
collisions in year 2007, we estimate for the full coverage the L0
trigger rate reduction for a single MTD hit will be 0.19 in
minimum-bias Au+Au collisions. Therefore online trigger
enhancement for di-muon trigger will reach 28. From 1 billion
minimum-bias Au+Au events, corresponding to 36 million di-muon
triggered events, we expect to get 5600 $J/\psi$ and 85 $\Upsilon$
through di-muon decays. Table~\ref{Tab:E} lists the L0 trigger
rate reduction for a single MTD hit with full MTD coverage in
different collision systems, which were estimated from the trigger
rates and enhancement factors for the prototype MTD tray in 200
GeV Au+Au collisions in year 2007 and d+Au and p+p collisions in
year 2008. Table~\ref{Tab:E} also indicates the trigger capability
in central Au+Au collisions for the di-muon program. In p+p
collisions it was found that when there was a coincidence with the
hit in the TOF detector in the same TPC sector, the trigger rate
was reduced by a factor of 3. From the extrapolation of a valid
MTD hit and the time difference measured between the VPDs on the
east and west sides of the interaction point, the expected hit
position in the TOF system was obtained. From the TOF system, the
module ID with a valid hit leads to another position measurement
since the modules in the TOF system are sequentially positioned
along the z direction. Shown in Fig.~\ref{ztofcorr} (left panel)
is the extrapolated z position versus measured module ID with a
valid hit in the TOF system in 200 GeV p+p collisions. Strong
correlations were observed. The difference of these two z values
are shown in the right panel of Fig.~\ref{ztofcorr}. A single
Gaussian was used to fit the distribution and the sigma was found
to be $\sim\!6$ cm. Combining the information in the MTD and TOF
system, we can obtain an order of magnitude rejection dramatically
increasing the efficiency for dimuon triggers in central Au+Au
collisions at 200 GeV.

\section{Physics perspectives with full coverage}\label{future}

In this section, we explore the physics potential of a full
coverage MTD for STAR at mid-rapidity. Two physics cases are
selected to illustrate the MTD capability for online triggering
and improvement of momentum resolution to achieve physics goals of
measuring $J/\psi$ elliptic flow and $R_{AA}$ at high-$p_T$ and
resolving ground state of $\Upsilon$ from its excited states. In
addition, electron-muon correlation can be used to distinguish
lepton pair production from heavy quark decays
($c+\bar{c}\rightarrow e+\mu(e)$, $B \rightarrow e(\mu)+ c
\rightarrow e+\mu(e)$).

\begin{figure}
\begin{center}
\includegraphics[keepaspectratio,scale=0.7]{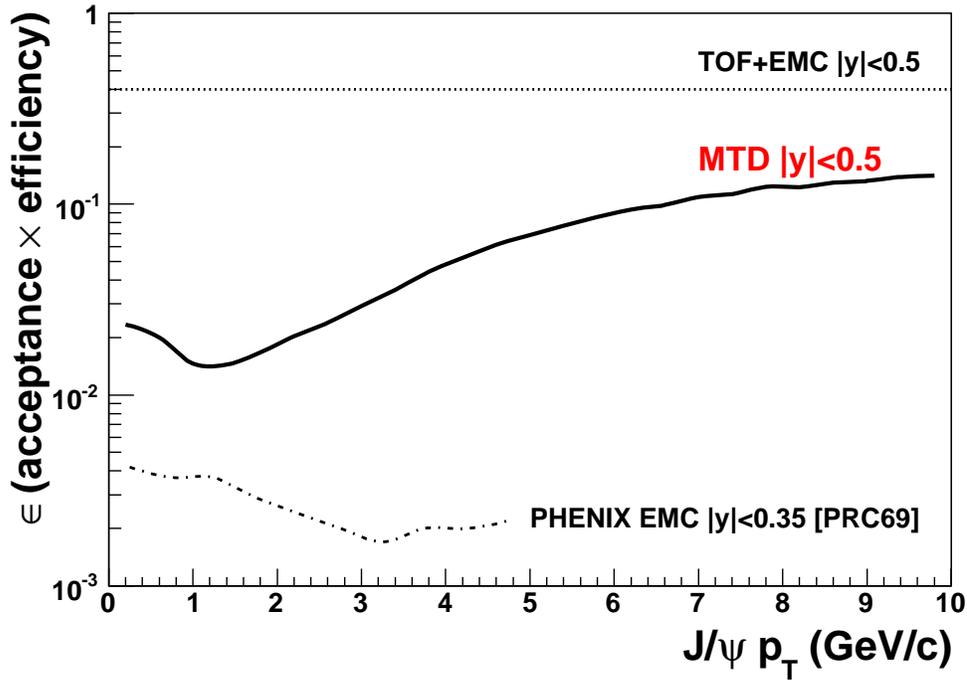}
\caption{Efficiency for $J/\psi$ as a function of $p_{T}$.
PHENIX~\cite{phenixjpsi} and STAR detect $J/\psi\rightarrow
e^+e^-$ at mid-rapidity using a combination of several detectors
together with electromagnetic calorimeters.} \label{figjpsieff}
\end{center}\end{figure}

Shown in Fig.~\ref{figjpsieff} are the efficiencies for $J/\psi$
at mid-rapidity at RHIC.  Both PHENIX and STAR are able to detect
$J/\psi$ at mid-rapidity through $J/\psi\rightarrow e^+e^-$ with
electromagnetic calorimeters as trigger device and electron
identifier. However, this is limited by the capability to trigger
on electrons at low momentum in STAR and relatively lower
efficiency times acceptance ($\epsilon$) in
PHENIX~\cite{phenixjpsi}. The proposed MTD detecting
$J/\psi\rightarrow \mu^+\mu^-$ will have much higher trigger
rejection power over the STAR EMC and TOF combination and have
larger acceptance over PHENIX configuration. This results in much
larger $J/\psi$ sample than the currently available RHIC
experimental setups.

\begin{figure}
\begin{center}
\includegraphics[keepaspectratio,scale=0.7]{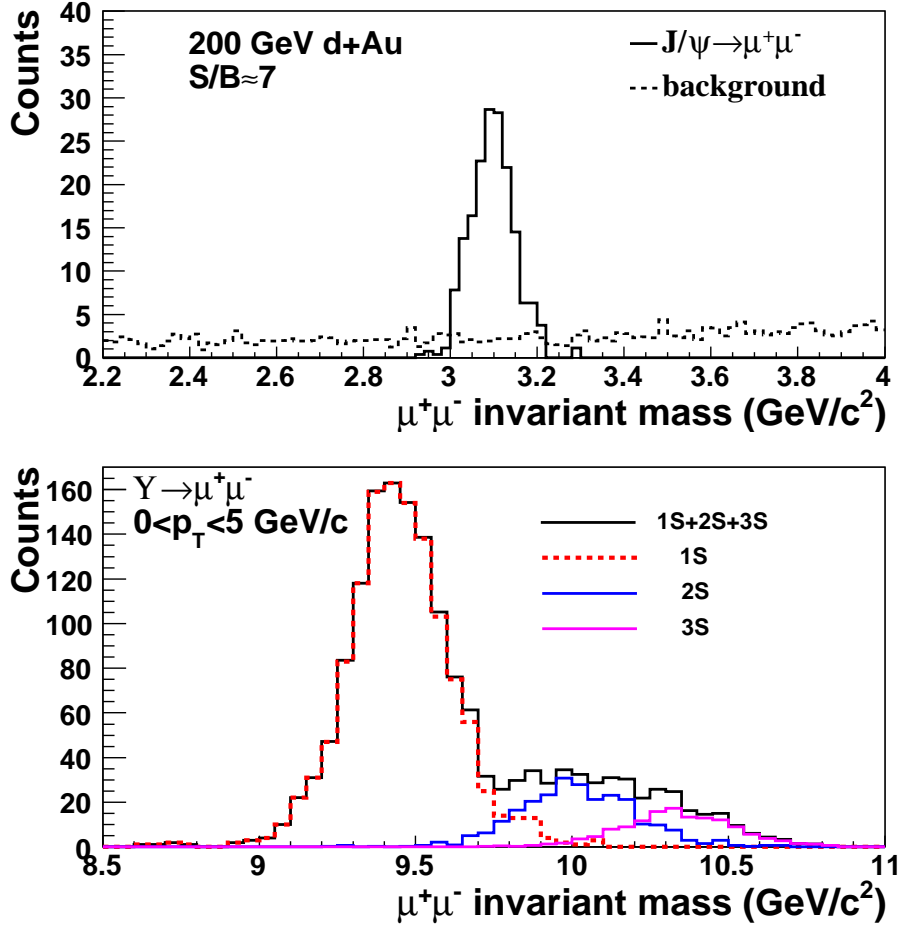}
\caption{ The expected di-muon invariant mass distribution from
$J/\psi$ and background in d+Au collisions (top panel); the
invariant mass distribution of di-muon decayed from $\Upsilon$ at
$0\!<p_T\!<\!5$ GeV/$c$ (bottom panel). The different $\Upsilon$
states can be separated. The background at $\Upsilon$ mass range
is not simulated since our small acceptance prototype has not
provided high enough $p_T$ reach.} \label{figureupsilon}
\end{center}\end{figure}

\begin{figure}
\begin{center}
\includegraphics[keepaspectratio,scale=0.7]{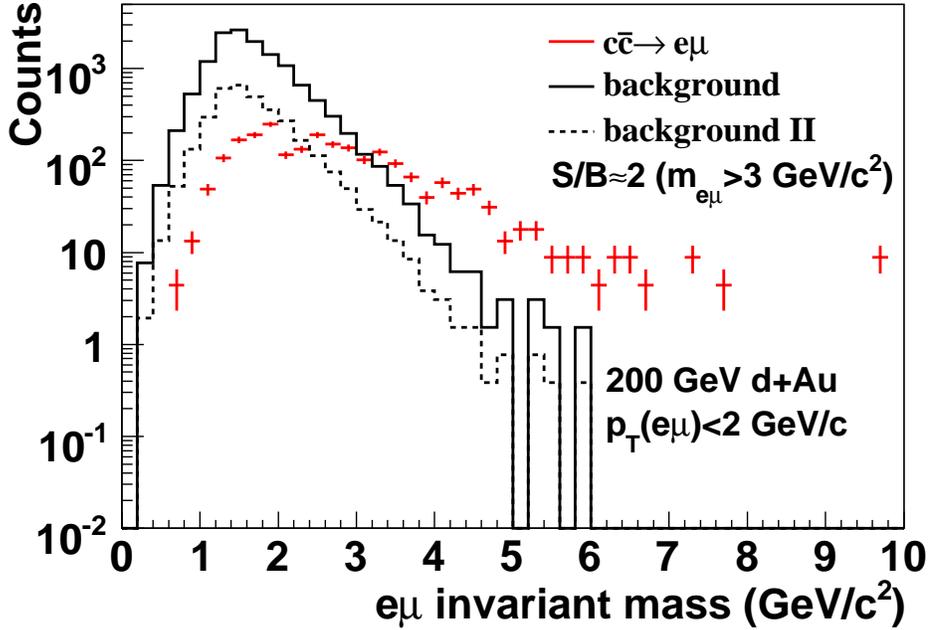}
\caption{ The expected electron-muon correlation from charm pair
production and from random background in d+Au collisions. The
statistics shown are equavelant to 2 billion minbias d+Au
collisions.  } \label{figureemuon}
\end{center}\end{figure}

Shown in Fig.~\ref{figureupsilon} (top panel) is the
$J/\psi\rightarrow\mu^{+}\mu^{-}$ with a signal-to-background
ratio of 7:1. The background di-muon pairs are simulated from the
inclusive muon yields obtained from studies shown in previous
sections. Shown in Fig.~\ref{figureupsilon} (bottom panel) is the
invariant mass distribution of di-muons decayed from Upsilon,
simulated in the STAR geometry. Clearly, different Upsilon states
($\Upsilon(1S)$, $\Upsilon(2S)$ and $\Upsilon(3S)$) can be
separated through the di-muon decay channel while Bremsstrahlung
energy losses of electrons present a challenge for the separation
due to the detector materials with future inner tracker upgrades
at STAR. Shown in Fig.~\ref{figureemuon} are the simulation
results of the electron muon invariant mass distribution from
charm pair production and from background. The cross symbols
represent the signal. The solid line represents the background,
obtained using random combinations between the photonic electron
yields measured in d+Au collisions~\cite{FuSQM08} and inclusive
muon yields shown in previous sections. The signal-to-background
ratio for charm pair production is 2:1 if the invariant mass of
the electron muon pair is larger than 3 GeV/$c^2$ and $p_T$ is
less than 2 GeV/$c$. The photonic electrons can be paired with
other particles in the TPC, which will lead to a factor of 2
rejection. The inclusive muons can be associated with the TOF
hits, resulting in at least another factor of 2 rejection. If we
include these rejection factors, the background will be reduced by
a factor of 4, as shown by the dashed line. The
signal-to-background ratio for electron muon correlation is
significantly enhanced.

In the RHIC II era, when 26 $nb^{-1}$ and 300 $pb^{-1}$ luminosity
for Au+Au and p+p collisions is delivered, we expect to get 169 k
$J/\psi$ and 2500 $\Upsilon$ in p+p collisions, and 630 k $J/\psi$
and 9300 $\Upsilon$ in Au+Au collisions~\cite{rhicIIQuarkonia}.
These measurements are essential to advance our understanding of
the QCD matter created at RHIC.

\section{Conclusions}\label{concl}
In summary, we propose a large-area, cost-effective muon telescope
detector for STAR and for the next generation of detectors at a
possible electron-ion collider. Cosmic ray and beam tests show the
intrinsic timing resolution of the long-MRPC is about 60-70 ps and
spatial resolution is better than 1 cm. The prototype of the MTD
tray triggered the data acquisition system at STAR successfully.
The results showed that offline tracking of particles from the TPC
was able to match hits from the long-MRPC. A muon purity of
$>\!80$\% can be achieved. The ratio of primary muon over
secondary muon was studied and found to be high quality for the
quarkonium program at RHIC.

\section{Acknowledgments}
We thank the STAR Collaboration and the RCF at BNL for their
support. This project is supported by the BNL Laboratory directed
R\&D fund 07-007. Long-MRPC R$\&$D at USTC is supported by
National Natural Science Foundation of China (10775131). L. Ruan
thank the Battelle Memorial Institute and Stony Brook University
for the support in the form of the Gertrude and Maurice Goldhaber
Distinguished Fellowship. Z. Xu is supported in part by the PECASE
Award.

\section*{References}

\end{document}